\documentclass[twocolumn]{jpsj3}
\usepackage{txfonts}

\title{Observation of Phonon-Assisted Magnon Absorption in Spin-Orbit Coupling Induced Mott Insulator Sr$_{2}$IrO$_{4}$}

\author{\name{Yasuyuki \surname{Hirata}}, \name{Hiroyuki \surname{Tajima}}, and \name{Kenya \surname{Ohgushi}}}
\inst{Institute for Solid State Physics, University of Tokyo, Kashiwanoha 5-1-5, Kashiwa, Chiba 277-8581, Japan} %\\

\kword{Infrared spectroscopy, Two-dimensional antiferromagnet, Mott insulator, Iridium oxides}

\begin{document}
\maketitle

The strong electron correlation effect in transition metal oxides is a rich source of various intriguing physical properties such as magnetism and superconductivity. One of the most imposing example is copper oxides such as La$_{2}$CuO$_{4}$ with the layered perovskite structure, where the high-temperature superconductivity emerges when carriers are doped into the antiferromagnetic Mott insulating state.\cite{hightc} Such exotic properties have typically been investigated with the respect to the 3$d$ electron system, because the electron correlation effect in the 4$d$/5$d$ electron systems is much weaker due to the more extended character of the wavefunctions. Nevertheless, recent studies have revealed \cite{kim1} that the Sr$_{2}$IrO$_{4}$ layered-perovskite with (5$d)^{5}$ valence electrons is a Mott insulator that undergoes the magnetic transition into a checkerboard-type spin arrangement at a N\'{e}el temperature of $T_{\rm N}$ = 240 K.\cite{crawford} This insulating state can be understood as follows: the strong spin-orbit interaction split the energy levels of the Ir 5$d$ $t_{2g}$ orbitals into doubly-degenerated $J_{\rm eff}$ = 1/2 and quadruply-degenerated $J_{\rm eff}$ = 3/2 orbitals. The $J_{\rm eff}$ = 1/2 orbitals form a half-filled band with a narrow bandwidth, which results in the enhanced electron correlation effect that leads to a Mott insulating state. A few other iridium oxides such as Ba$_{2}$IrO$_{4}$ and CaIrO$_{3}$ are also magnetic insulators.\cite{ba2iro4,cairo3} Carriers can be doped by chemical substitution in these spin-orbit coupling induced Mott insulators, so that they are potential high-temperature superconductor candidates.\cite{irsc,subst}

To clarify electronic states and pursuit the possibility of high-temperature superconductivity in Sr$_2$IrO$_4$ requires the information on the superexchange interaction between two Ir spins $J$. However, in contrast to the well-studied on-site character of Ir orbitals,\cite{kim2,moon} there have been few studies on the inter-site interactions. Jackeli and Khaliullin theoretically calculated $J$ to be 45 meV,\cite{jackeli} while $J$ was estimated to be 60 meV from detection of the single-magnon dispersion with resonant inelastic x-ray scattering.\cite{rixs} To establish a reliable experimental evaluation of $J$, utilizing another method is highly desired. Infrared transmission measurement is a powerful technique for the detection of magnons.\cite{perkins1,perkins2,lorenzana} Although single magnon excitations are optically forbidden under the spatial inversion symmetry, two magnons coupled with one phonon can be excited by light, which is detected as a phonon-assisted magnon absorption. $J$ values have been successfully estimated from the peak energy of phonon-assisted magnon absorption in the absorption spectrum for La$_{2}$NiO$_{4}$ and La$_{2}$CuO$_{4}$.\cite{perkins1,perkins2,lorenzana}

In this paper, we report on magnon excitation in the antiferromagnetic Mott insulator Sr$_{2}$IrO$_{4}$, as revealed by optical transmission measurements. We have successfully observed phonon-assisted magnon absorption in the absorption spectrum and deduced $J$ = 57 meV, which is consistent with the resonant inelastic x-ray scattering measurement.

Plate-like single crystals of Sr$_{2}$IrO$_{4}$ (1$\times$1$\times$0.2 mm$^3$) were grown using the flux method.\cite{kim2} SrCO$_{3}$, IrO$_{2}$ and SrCl$_{2}$ powders were mixed in a molar ratio of 3:1:15, heated to 1300 $^{\circ}$C, and then cooled to 900 $^{\circ}$C at a cooling rate of 8 $^{\circ}$C/h. The $ab$-planes of the obtained crystals were polished with Al$_{2}$O$_{3}$ powders until the thickness became 5-10 $\mu$m. Transmission spectra were measured using a Fourier-transform infrared spectrometer with the geometry of incident light along the $c$-axis and polarization along the $a$-axis in the frequency range between 600 and 4000 cm$^{-1}$. The temperature was controlled in the range of 20 and 300 K using a He flow cryostat.

\begin{figure}
\includegraphics[width=8cm]{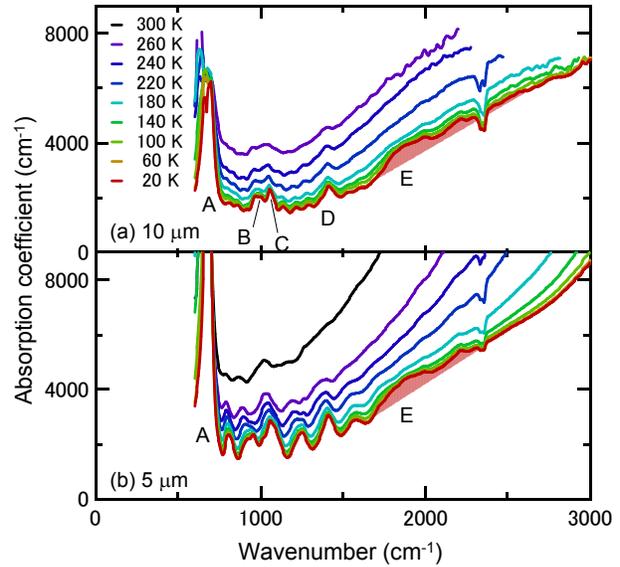}%
\caption{\label{fig:spectra}Absorption spectra as a function of photon energy for Sr$_{2}$IrO$_{4}$ with thicknesses of (a) 10 and (b) 5 $\mu$m. The peak structure labeled as A is the oxygen stretching phonon mode, and those labeled as B, C, and D are two-phonon modes. The broad structure labeled as E (shaded area) is the phonon-assisted magnon absorption (see text for details). The dip structure around 2400 cm$^{-1}$ is an artifact originating from CO$_{2}$ absorption.}
\end{figure}

Figure \ref{fig:spectra} shows absorption spectra of Sr$_{2}$IrO$_{4}$ with thicknesses of (a) 10 and (b) 5 $\mu$m. The background absorption increases linearly with the photon energy and originates from the low-energy tail structure of the Mott excitation centered around 0.54 eV,\cite{moon} and decreases monotonically on cooling. The spectra for the 10 $\mu$m sample [Fig.~\ref{fig:spectra}(a)], has an artifact originating from CO$_{2}$ absorption at around 2400 cm$^{-1}$, and peak structures at 698 (A), 972 (B), 1042 (C), and 1415 cm$^{-1}$ (D) are discernible at all measured temperatures. In addition, a broad peak appears at 1870 cm$^{-1}$ (E) only in the spectra measured below 180 K. In the spectra for the 5 $\mu$m sample [Fig.~\ref{fig:spectra}(b)], the B, C, and D modes are concealed by the oscillating features due to the interference of backward reflections; however, the A and E peak structures can be recognized more clearly in the thinner sample. The origin of these features can be elucidated with reference to phonon frequencies determined from reflectivity measurements.\cite{moon} Sr$_{2}$IrO$_{4}$ has three in-plane optical phonons that involve oxygen vibrations; two bending modes at 284 and 356 cm$^{-1}$, and one stretching mode at 663 cm$^{-1}$.\cite{moon} Peak A can therefore be reasonably assigned to the oxygen stretching mode. Peaks B, C, and D are interpreted as two-phonon modes; 284 and 663 cm$^{-1}$ phonons for peak B, 356 and 663 cm$^{-1}$ phonons for peak C, and two 663 cm$^{-1}$ phonons for peak D.

The broad E structure centered at 1870 cm$^{-1}$ cannot be explained by either one-phonon or two-phonon excitations. This mode is developed only below 180 K, which is lower than $T_{\rm N}$ = 240 K, which suggests a magnetic origin. Moreover, the peak width of {\it ca.} 200 cm$^{-1}$ and the peak height of {\it ca.} 500 cm$^{-1}$ are comparable with those of the phonon-assisted magnon absorption in La$_{2}$CuO$_{4}$, where the peak width is {\it ca.} 300 cm$^{-1}$ and the peak height is {\it ca.} 200 cm$^{-1}$.\cite{perkins1} The high-energy tail structure observed in the phonon-assisted magnon absorption of La$_{2}$CuO$_{4}$ is not explicitly recognized in Sr$_{2}$IrO$_{4}$. This is most likely because the tail structure laps over the large background electronic absorptions; we here assumed a simple extrapolation for the tail structure and indicated the magnon contributions by the shaded area in Fig.~\ref{fig:spectra}. Considering these issues, we conclude that the broad E structure is the phonon-assisted magnon absorption, where two magnons and one phonon are excited.

According to the theoretical study by Lorenzana and Sawatzky, if the nearest-neighbor $S$ = 1/2 Heisenberg Hamiltonian is considered, then the peak energy of the phonon-assisted magnon absorption is represented as $E_{\rm infrared} = 2.73 J + \omega_{\rm ph}$, where $\omega_{\rm ph}$ is the energy of the assisting phonon.\cite{lorenzana} If we suppose that the 663 cm$^{-1}$ oxygen stretching mode involves the observed phonon-assisted magnon mode, as in the case of La$_{2}$CuO$_{4}$, then $J$ is calculated to be 57 meV.\cite{bending} This is fairly close to the estimation from resonant inelastic x-ray scattering ($J$ = 60 meV), although direct comparison is not appropriate because not only the nearest-neighbor exchange interaction $J$, but also the next-nearest-neighbor exchange interaction $J'$, and the third-nearest-neighbor exchange interaction $J''$, are considered in the latter estimation.\cite{rixs} Cetin {\it et al.} reported a broad mode around 1800 cm$^{-1}$ in the Raman spectra. The mode is developed below $T_{\rm N}$ and is assigned to a two-magnon excitation.\cite{raman} In the framework of the nearest-neighbor $S$ = 1/2 Heisenberg model, the peak energy of the two-magnon mode is represented as $E_{\rm Raman} = 3.38 J$;\cite{canali} therefore, the exchange interaction is estimated to be $J$ = 66 meV, which is consistent with our present result.  On the other hand, the results of the resonant magnetic x-ray diffuse scattering give  $J \sim$ 0.1 eV, which is larger than our conclusion.\cite{fujiyama} The reason of this discrepancy is not clear at present.

$J$ = 57 meV for Sr$_{2}$IrO$_{4}$ is approximately half that of the reported $J$ = 121 meV for La$_{2}$CuO$_{4}.$\cite{lorenzana} This indicates that the energy scale of Mott physics in iridium oxides would be rather large, even in a $5d$ system;\cite{irsc} if it is simply assumed that the superconducting transition temperature $T_{c}$ is proportional to $J$, then the present result suggests that doped iridium oxides can be potential superconductors with $T_{c}$ comparable to high-$T_{c}$ copper oxide superconductors.

In conclusion, infrared transmission spectroscopy was performed to investigate Sr$_{2}$IrO$_{4}$ as a spin-orbit coupling induced Mott insulator. A phonon-assisted magnon absorption with a peak energy of 1870 cm$^{-1}$ is developed below 180 K. The nearest-neighbor superexchange interaction is estimated to be $J$ = 57 meV, which is consistent with results obtained from resonant inelastic x-ray scattering and Raman scattering measurements. The $J$ value is approximately half that of $J$ for La$_{2}$CuO$_{4}$, which suggests that a novel phenomenon is realized at rather high temperature, even in 5$d$ transition metal oxides.

We are grateful to Y. Ueda, M. Isobe and S. Fujiyama for helpful discussions and experimental support. This work was supported by Special Coordination Funds for Promoting Science and Technology, Promotion of Environmental Improvement for Independence of Young Researchers, and a Grant-in-Aid for Scientific Research (B) (No. 20740211).


\begin{thebibliography}{99}
\bibitem{bending}
If the assisting phonon is 284 or 356 cm$^{-1}$ mode, then the estimated $J$ value becomes 74 or 71 meV, respectively.
\end{thebibliography}
\end{document}